\documentclass[letter]{aa} 

\usepackage{graphicx}
\usepackage{amssymb}	
\usepackage{mathtools}
\usepackage{gensymb}
\usepackage{amsmath}
\usepackage[colorlinks=true,linkcolor=blue,allcolors=blue]{hyperref}%

\usepackage{txfonts}
\def\ltsima{$\, \buildrel < \over \sim \,$}
\def\simlt{\lower.5ex\hbox{\ltsima}}
\def\gtsima{$\, \buildrel > \over \sim \,$}
\def\simgt{\lower.5ex\hbox{\gtsima}}

\begin{document}

   \title{New KiDS in town}

   \subtitle{Sextans~II: a new stellar system in the outskirts of the Milky Way}

   \author{Massimiliano Gatto
          \inst{1}\fnmsep\thanks{massimiliano.gatto@inaf.it}
          \and
          Michele Bellazzini\inst{2}
          \and
          Crescenzo Tortora\inst{1}
          \and
           Vincenzo Ripepi\inst{1}
          \and
          Massimo Dall'Ora\inst{1}
          \and
          Michele Cignoni\inst{1,3,4}
          \and
          Konrad Kuijken\inst{5}
          \and
          Hendrik Hildebrandt\inst{6}
          \and
          Shiyang Zhang\inst{6}
          \and
          Jelte de Jong\inst{5,7}
          \and
          Nicola R. Napolitano\inst{8}
          \and Simon E. T. Smith\inst{9}
          }

   \institute{INAF-Osservatorio Astronomico di Capodimonte, Via Moiariello, 16, I-80131, Napoli, Italy
         \and
   INAF-Osservatorio di Astrofisica e Scienza dello Spazio, Via Gobetti, 93/3, I-40129, Bologna, Italy  
   \and
   Physics Department, University of Pisa, Largo Bruno Pontecorvo 3, I-56127, Pisa, Italy
   \and
   INFN-Largo Bruno Pontecorvo 3, I-56127, Pisa, Italy
   \and
   Leiden Observatory, Leiden University, P.O.Box 9513, 2300RA Leiden, The Netherlands
   \and
   Ruhr University Bochum, Faculty of Physics and Astronomy, Astronomical Institute (AIRUB), German Centre for Cosmological Lensing, 44780 Bochum, Germany
   \and
   Kapteyn Astronomical Institute, University of Groningen, PO Box 800, 9700 AV Groningen, The Netherlands
   \and
   School of Physics and Astronomy, Sun Yat-sen University, Guangzhou 519082, Zhuhai Campus, P.R. China
   \and Department of Physics and Astronomy, University of Victoria, Victoria, BC, V8P 1A1, Canada
             \\
             }

   \date{}

 
  \abstract

\abstract{We report on the discovery of a significant and compact over-density of old and metal-poor stars in the KiDS survey (data release 4). The discovery is confirmed by deeper HSC-SSC data revealing the old Main Sequence Turn-Off of a stellar system located at a distance from the sun of $D_{\sun}=145^{+14}_{-13}$~kpc in the direction of the Sextans constellation. The system has absolute integrated magnitude ($M_V=-3.9 \pm 0.4$), half-light radius ($r_h=193^{+61}_{-46}$~pc), and ellipticity ($e=0.46^{+0.11}_{-0.15}$) typical of Ultra Faint Dwarf galaxies (UFDs). The central surface brightness is near the lower limits of known local dwarf galaxies of similar integrated luminosity, as expected for stellar systems that escaped detection until now. The distance of the newly found system suggests that it is likely a satellite of our own Milky Way, consequently, we tentatively baptise it Sextans~II (KiDS-UFD-1).}


   \maketitle
%

\section{Introduction}

The past two decades have seen a surge of interest in the research of Milky Way (MW) satellites, thanks to the advent of digital wide-field deep panchromatic photometric surveys.
The application of high-performance over-density detection techniques to large data sets has led to a substantial increase in the number of known stellar systems inhabiting the MW halo \citep[e.g.][and references therein]{Belokurov-2007,Walsh-2009,Koposov-2015,Torrealba-2016,Cerny-2021}. The quest for MW satellite galaxies carried out with the modern facilities opened up the study of the faint end of the galaxy luminosity function down to
the so-called ultra-faint dwarf (UFD) galaxies \citep[see][for discussion and references]{Belokurov-2013}. 
These objects provide valuable insights into the mass assembly history of the MW. In addition, the UFDs are the most dark-matter-dominated objects in the Universe, as well as 
the faintest, oldest and least chemically evolved galaxies known so far \citep[see for example the detailed review by][and references therein]{Simon2019}. Therefore, they represent fossil evidence to probe the very early stages of the Universe, such as the epoch of the reionization \citep[e.g.][]{Bovill&Ricotti2011,Wheeler-2015}. Because of their pristine nature, UFDs are also ideal systems for testing and improving sub-branches of the stellar evolutionary models, such as the synthesis of heavy elements \citep[e.g.][]{Ji-2016}.\par
The majority of faint galaxies discovered in recent years were identified through a few large sky surveys that offer unprecedented photometric depth, including the Sloan Digital Sky Survey \citep[SDSS;][]{Willman-2005a, Willman-2005b, Belokurov-2006b,Belokurov-2007,Belokurov-2009,Belokurov-2010,Zucker-2006a,Zucker-2006b}, the Dark Energy Survey \citep[DES;][]{Bechtol-2015,Drlica-Wagner-2015,Kim&Jerjen2015,Koposov-2015,Luque-2016}, PanSTARRS \citep[][]{Laevens-2015a,Laevens-2015b}, ATLAS \citep{Torrealba-2016}, DELVE \citep{Mau-2020,Cerny-2021,Cerny-2023}, DESI \citep{Martinez-Delgado-2022,Sand-2022}, MagLiteS \citep{Torrealba-2018}, HSC-SSP \citep{Homma-2016,Homma-2018,Homma-2019}, and UNIONS \citep{Smith-2023a,Smith-2023b}.
In this study, we expand the search for unknown stellar systems to the Kilo-Degree Survey \citep[KiDS;][]{deJong-2013}, which has not been exploited yet for this purpose.\par
KiDS is an extensive multi-band photometric survey conducted using the VLT Survey Telescope \citep[VST;][]{Capaccioli&Schipani2011,Capaccioli-2012}. 
KiDS has observed about 1350 square degrees of the sky in the $u, g, r$, and $i$ bands. KiDS was conceived as a cosmological survey, mainly to use weak gravitational lensing of galaxies to study the assembly of large-scale structures. However, as a side-product of extra-galactic studies, KiDS provides a deep and accurate catalogue of stars, making it suitable to search for low-surface brightness stellar systems in the Local Group.
In this letter, we present a new UFD galaxy uncovered in a large-scale search for low-luminosity stellar overdensities over the whole KiDS Data Release 4 area.

\section{KiDS Data}
\label{sec:data}

We use the most recent KiDS data release \citep[DR4,][]{Kuijken-2019}, which includes 1006 of imaging and catalogue data
with limiting $5\sigma$ AB magnitudes of $24.23 \pm 0.12$, $25.12 \pm 0.14$, $25.02 \pm 0.13$ and $23.68 \pm 0.27$ mag in the $u$, $g$, $r$, and $i$ filters, respectively, that is $\ga 2$ mag deeper than SDSS and PanSTARRS DR1. 
DR4 also includes photometry in the $Z, Y, J, H, K_s$ infrared (IR) bands obtained with the VISTA Kilo-degree INfrared Galaxy survey \citep[VIKING;][]{Edge-2013} carried out on the VISTA telescope.
KiDS covers two different stripes in the sky, a Northern and a Southern stripe \citep[KiDS-N and KiDS-S, hereafter, see Table 2 in][for the exact boundaries]{Kuijken-2019}.\par 
From the DR4 catalogue, we selected sources identified as stars, namely objects that are flagged 1, 4 or 5 in the {\sc SG2DPHOT} column of the catalogue \citep[see A.1.2 in][]{Kuijken-2019}. In this way, we obtained a list of 4,422,730 stars in KiDS-N and 3,116,050 stars in KiDS-S\footnote{Magnitudes provided by KiDS are corrected for interstellar extinction using the \citet{Schlegel-1998} reddening maps re-calibrated after \citet{Schlafly&Finkbeiner2011}, in combination with the Fitzpatrick's law \citep{Fitzpatrick-1999}}.
Successively, we filtered out sources with low-quality photometry by imposing a maximum tolerated photometric uncertainty in both $g$ and $r$ bands of 0.2 mag.
Nonetheless, the catalogues at our disposal still include distant galaxies that have been erroneously classified as stars. To enhance the purity of our catalogues, we took advantage of the multi-band photometry provided by KiDS, utilizing colour-colour diagrams to further purge unresolved galaxies.
Specifically, we focused on the $g-r$ versus $g-i$ plot, where stars occupy a distinctive, narrow sequence \citep[see for example Fig.~18 in][]{Kuijken-2019}. 
To this purpose, we selected only sources falling within the boundaries of a polygon defined by the following coordinates (expressed in mag): $(g-r, r-i) = (1.25, 0.25); (1.25, 0.8); (-0.65, -0.05); (-0.65, -0.6)$. It is important to note that we also opted to exclude red stars, characterized by $g-r > 1.25$ mag, as this colour range is dominated by M-type dwarf stars located within the MW disk. After the colour-colour selection we are left with 3,558,843 stars, 2,127,949 of which are located within the footprint of KiDS-N and the remaining 1,430,894 stars in KiDS-S.\par

\section{Method}

We tackled the task of unveiling unknown faint MW satellites by identifying the over-densities of stars in the sky relative to the local average stellar background. To achieve this, we utilized the algorithm and the procedure described in \citet{Gatto-2020}.
In addition, to augment the chances of detecting faint stellar systems projected against the dominant MW stellar population, we also followed the approach of \citet{Walsh-2009}, which has been widely used in recent years for the discovery of low-luminosity galaxies and globular clusters in the Milky Way halo \citep[e.g.][and references therein]{Bechtol-2015,Koposov-2015,Torrealba-2016,Homma-2019,Cerny-2021}.
This method consists of a preliminary isochrone-based matched filter on the catalogue of stars.
In particular, prior to running the searching algorithm, we selected stars whose positions in the $g-r, r$ colour-magnitude diagram (CMD) were within 0.05 mag in colour from a given isochrone representative of an old and metal-poor stellar population\footnote{We took into account also photometric uncertainties of the stars to decide whether or not a star is inside to the isochrone filter.}, which is characteristic of the faint galaxies expected to inhabit the MW halo. 
We opted to work with the suite of PARSEC isochrones \citep{Bressan-2012} and the isochrone filter was carried out by adopting three different ages, namely $\log (t/yrs)$ = 10.0, 10.06, 10.12 dex and three different metallicities, or [Fe/H] $\sim$ -2.2, -1.5, -1.2 dex.
Additionally, we varied the distance modulus of the isochrone in a range between 16 and 23.6 mag (corresponding to a range of physical distances between $\sim$16 and $\sim$520 kpc) with steps of 0.4 mag. Therefore, we scanned the KiDS catalogue with 180 different isochrone filters.\par
For each isochrone, we analyse the resulting filtered map as follows. We divide the map into 4\degree $\times$ 4\degree~regions, and then build density maps using the star coordinates as input parameters,  moothed through a Kernel Density Estimation (KDE) technique adopting a Gaussian kernel.
Subsequently, the algorithm counts the number of stars ($N_{\rm stars}$) in each pixel of the sub-field and measures the number of standard deviations above a local mean:
\begin{equation}
    S_{\rm pxl} = (N_{\rm stars} - N_{\rm bkg})/\sigma_{\rm bkg}
\end{equation}

\noindent
where $N_{\rm bkg}$ and $\sigma_{\rm bkg}$ represent the average of the local stellar background and its standard deviation, respectively\footnote{Please refer to \citet{Gatto-2020}, in particular their Sect.~3.1, for a detailed description of the detection algorithm and definitions of all the involved quantities.}.
In this work, we used a pixel size of 30\arcsec $\times$ 30\arcsec, and a bandwidth of the Gaussian kernel function of 1\arcmin\footnote{Background quantities were estimated in a box window between 10\arcmin~and 20\arcmin~from the pixel.}.
As a sanity check, we conducted a validation of the algorithm by assessing its capability to accurately identify known stellar systems residing within the KiDS footprint, including Galactic globular clusters and dwarf galaxies. In this way, our algorithm recovered Leo~V with $S_{\rm pxl}= 21.5$, by applying an isochrone filter\footnote{It is worth noting that the algorithm 
may detect a given over-densities with more than one reference isochrone. In these case we always consider the detection yielding the highest $S_{\rm pxl}$.} with $\log t$ = 10.06 dex, [Fe/H] $\sim$ -2.2 and a distance modulus = 21.2 mag, corresponding to a distance of approximately 175 kpc. This determination closely aligns with the previously estimated distance to the dwarf, which is reported as $173 \pm 5$ kpc by \citet{Medina-2017}.\par

\section{Results}

Utilizing the methodology outlined in the previous section, we identified a highly promising over-density in the Sextans constellation. 
Our detection revealed a $S_{\rm pxl}=10.6$  stellar over-density by applying an isochrone filter with $\log (t/yrs)$ = 10.12 dex, [Fe/H] $\sim$ -1.5, and a distance modulus $(m - M)_0$ = 20.4 magnitudes.
The left panel of Figure 1 shows the density map of stars in the region of the newfound galaxy candidate, for the aforementioned isochrone selection filter. The overdensity stands out clearly.
The subsequent three panels of the figure illustrate the CMDs of stars within a 5\arcmin~radius from the over-density's centre, the CMD of a representative local field covering an equivalent area, and the Hess diagram derived through the subtraction of the over-density and field CMDs, from left to right, respectively.
\begin{figure*}
    \centering
    \includegraphics[width=0.35\textwidth]{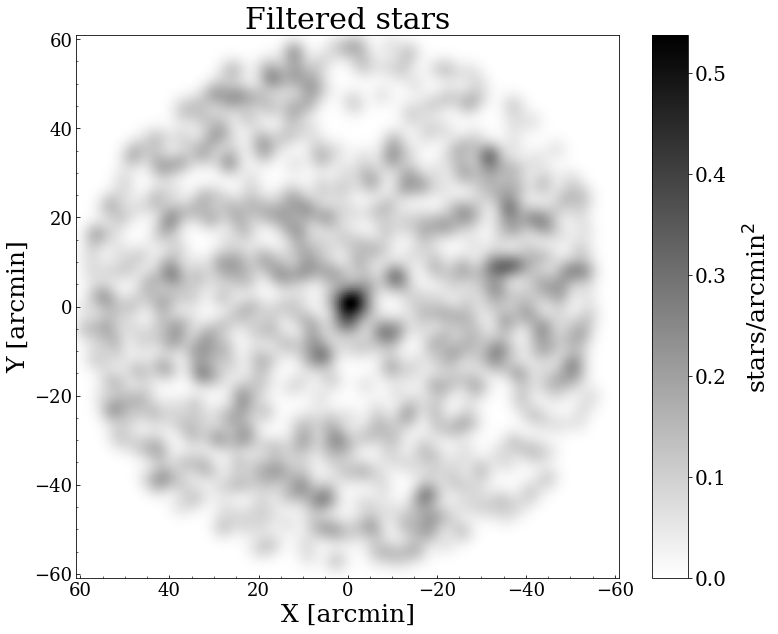}
    \includegraphics[width=0.55\textwidth]{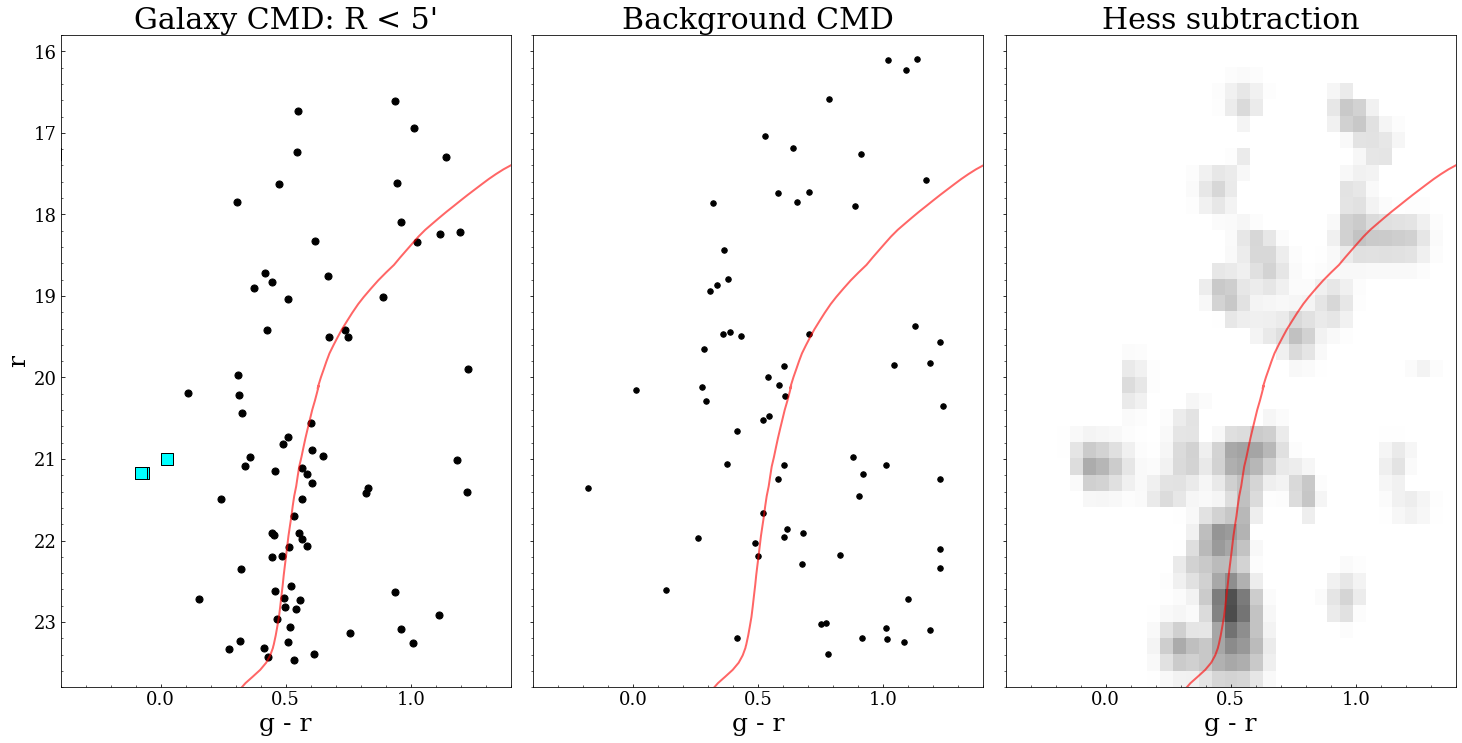}
    \caption{\emph{First panel}: Density map of a region with radius 1\degree~centered on the galaxy candidate. The density map is obtained using stars from the KiDS catalog, and filtered by an isochrone with $\log (t/yrs)$ = 10.12 dex, [Fe/H] $\sim$ -1.5 and a distance modulus = 20.4 mag. We adopted a Gaussian kernel with bandwidth = 2.0\arcmin~to smooth the figure. \emph{Second panel:} CMD of stars located within 5\arcmin~from the over-density center. A red isochrone with $\log (t/yrs) = 10.12$ dex; $[Fe/H] \sim$ -1.5 dex, and $m -M$ = 20.4 mag is superimposed on the figure. Cyan squares indicate the position of stars compatible with the HB evolutionary stage. \emph{Third panel}: CMD of stars within a local field having the same area covered by the CMD built for the dwarf galaxy candidate. \emph{Fourth panel}: Hess diagram obtained by subtracting the CMDs displayed in the second and third panels.}
    \label{fig:kids-1_detection}
\end{figure*}
The inspection of these CMDs unveils a well-sampled red giant branch (RGB) within the inner 5\arcmin, that is absent in the local field CMD, suggesting the possible detection of a stellar system. Three likely candidate horizontal branch (HB) stars, highlighted as cyan squares in the figure, are found at $(g-r, r) \simeq (0, 21-21.5)$ mag\footnote{Note that the PARSEC library do not include the zero age horizontal branch for the mentioned metallicity value.}.
Unfortunately, the photometric depth of the KiDS data is not sufficient to reach the main-sequence turn-off (MSTO) of the candidate\footnote{Although $5\sigma$ AB magnitudes reach $r \simeq$ 25 mag, the cuts applied to the catalog, described in Sect.~\ref{sec:data}, limited our analysis to $r \simeq 23.5 -24$ mag in the CMD, preventing us from probing fainter magnitudes.}, whose detection would be crucial to confirm it as a genuine stellar system and to assess its properties better. In the next section, we describe the successful follow-up of the candidate with deeper photometry.


\subsection{Deeper photometry in HSC-SSP}

The Hyper Suprime Cam Subaru Strategic Program \citep[HSC-SSP;][]{Aihara-2018a,Aihara-2018b} is a wide-field optical-near infrared survey conducted using the 8.2-meter Subaru Telescope. The HSC-SSP is designed to cover approximately 1400 square degrees of the Northern sky across five distinct filters, namely $g, r, i, z, y$. 
The survey 5$\sigma$~ magnitude limits for point-sources are impressively deep: 26.5 mag in the $g$ band, 26.1 mag in the $r$ band, and 25.9 mag in the $i$ band \citep[][]{Aihara-2018a}, surpassing the limiting magnitude of the KiDS survey by more than one mag in each passband. 
HSC-SSP has already demonstrated its power for the detection of low surface-brightness stellar systems \citep{Homma-2016,Homma-2018,Homma-2019}. In particular, \citet{Homma-2019} reported the discovery of Bootes~IV, one of the faintest UFDs identified to date, characterized by surface brightness levels $\mu \geq 32$~mag arcsec$^{-2}$. Since our candidate lies within the HSC-SSP footprint, this survey provides an excellent means to follow it up.\par
From the HSC-SSP catalogue provided by the public third data release \citep{Aihara-2022}, which included data for slightly less than 700 square degrees, we selected point-like sources within the region of the newly discovered over-density by means of the \emph{extendedness} parameter in $i-$band, which is the filter with the best median seeing \citep[i.e. $\sim$ 0.6\arcsec,][]{Aihara-2022}.  
To further refine the star-galaxy separation, we used a color-color diagram, as done above for the KiDS data. In this case, we adopted $(g - r,z - y)$ diagram that is particularly efficient in this sense.
Consequently, we selected all point-like sources located within the polygon defined by the criteria: $(g - r,z - y) = (-0.3,-0.12); (-0.3,0.08); (1.2,0.25); (1.2,0)$.
The following sections of this manuscript are based on the analysis of this catalogue.

\subsection{Analysis of the CMD}
\label{sec:cmd_fitting}

\begin{figure*}
    \centering
    \includegraphics[width=0.46\textwidth]{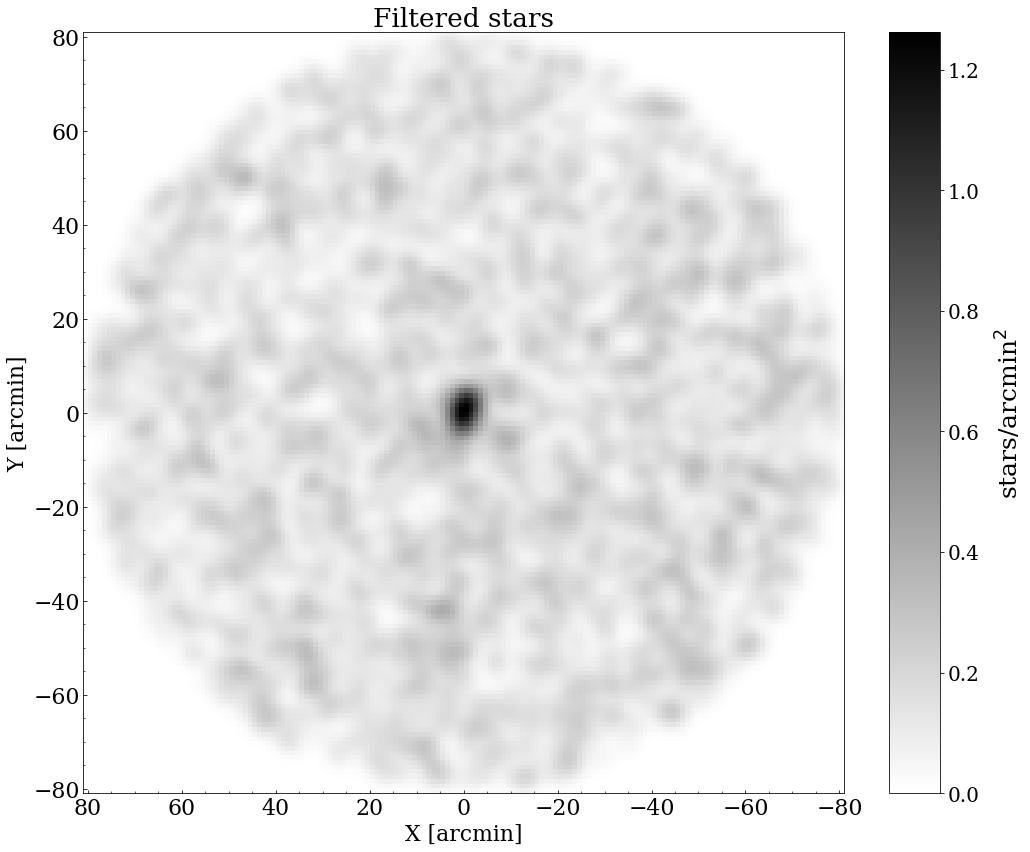}
    \includegraphics[width=0.46\textwidth]{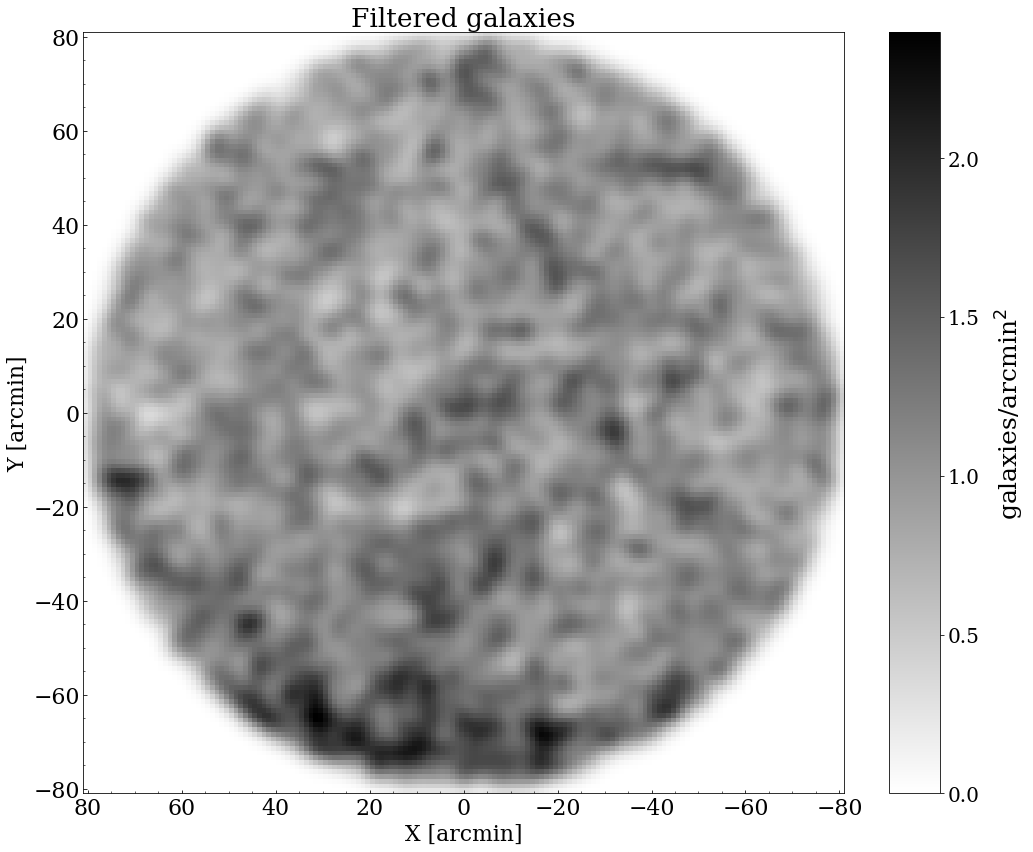}\\
    
    \includegraphics[width=0.9\textwidth]{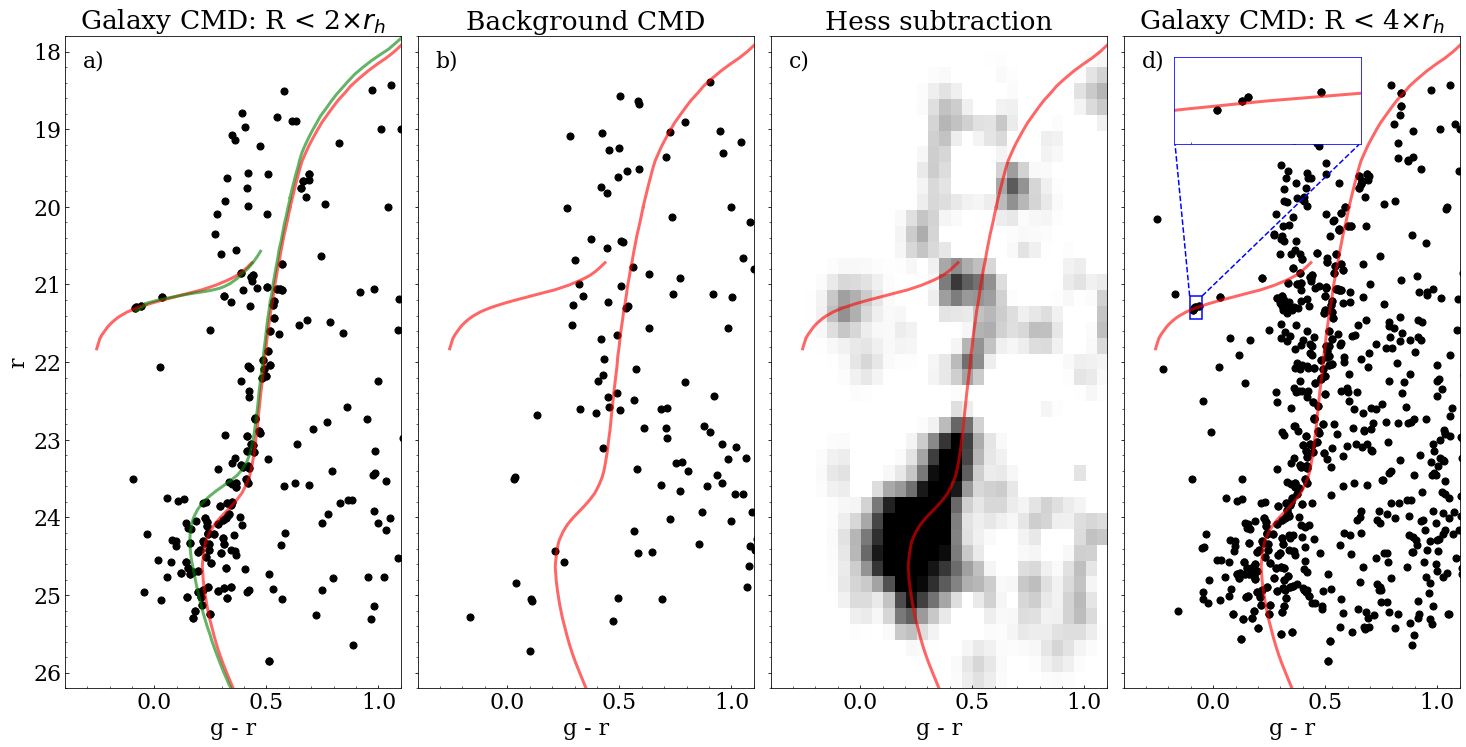}\\
    
    \caption{\emph{Top row - Left}: Density map of stars included in the HSC-SSP catalog, and filtered by an isochrone characterized by $\log (t/yrs) = 10.12$ dex; [Fe/H] $\sim$-2.2 dex, and $m -M$ = 20.8 mag. The depicted region encompasses 80\arcmin~in radius centered on the dwarf galaxy. The figure has been smoothed with a Gaussian kernel with bandwidth = 2\arcmin. \emph{Top row - Right}: Density map of sources included in the HSC-SSP catalog and classified as galaxies, filtered by the same isochrone parameters adopted for the stars. 
    \emph{Bottom row - panel a}: CMD of the stars located within $2 \times r_h$ from the galaxy center. A red (green) isochrone with $\log (t/yrs) = 10.00$ ($log (t/yrs) = 10.12$); [Fe/H] $\sim$-2.2 dex, and $m -M$ = 20.8 mag is overlaid to the figure. \emph{Bottom row - panel b}: CMD of stars within a local field of the same area as that used to construct the CMD in the panel to the left. \emph{Bottom row - panel c}: Hess diagram obtained by subtracting the CMDs displayed in the second and third panel. \emph{Bottom row - panel d}: CMD of the stars located within $4 \times r_h$ from the galaxy center. An inset positioned at the top of the panel offers a zoomed-in view of the HB stage, capturing a narrow color interval and highlighting the presence of four closely-aligned HB stars.}
    \label{fig:kids_density_maps}
\end{figure*}

In Fig.~\ref{fig:kids_density_maps} we show the results of the analysis of the deeper HSC-SSP catalogue.
The top left panel shows the density map of the stars filtered by the best-fit isochrone (indicated by a red solid line in the bottom row of the same figure), fully confirming the detection of a strong and compact stellar over-density at the expected location.
In contrast, the right panel of the same figure illustrates the density map for sources catalogued as galaxies based on the \emph{extendedness} parameter and subsequently filtered by the same isochrone. Notably, no discernible over-density of galaxies in the same region is observed. \\
The panel $a$ of the second row of Fig.~\ref{fig:kids_density_maps} displays the CMD\footnote{It is important to highlight that the HSC filter system is slightly different from the SDSS filter system. In order to work with the SDSS filter system available in the PARSEC library, we adopted the correction reported in \citet{Homma-2016}.} of all stars located within twice the half-light radius of the structure (for the determination of $r_h$, please refer to Section~\ref{sec:structural_params}).
The Sub Giant Branch (SGB) and the MSTO associated with the RGB already detected in KiDS data are unequivocally revealed. Notably, these features are not present in the CMD of the control field (same area, located 20\arcmin~from the centre of the over-density; see also the Hess diagram obtained by subtracting the two former CMDs). 
This compelling evidence strongly suggests that our candidate is a genuine stellar system, likely a faint local dwarf galaxy, hitherto unreported in the existing literature. In what follows we shall refer to it as Sextans~II (KiDS-UFD-1)\footnote{Concurrently and independently to our research, \citet{Homma-2023} also reported the discovery of Sextans~II by using the HSC-SSP catalogue.}.\par
An in-depth analysis of the CMD  also discloses the presence of four stars likely belonging to the HB evolutionary phase
Specifically, two stars with colours of -0.06 mag and -0.08 mag reside within the half-light radius of the galaxy. The third star, slightly exceeding the $r_h$, has $g - r$ = -0.09 mag, while the fourth one has $g - r$ = 0.03 and is located at  $\sim$ 8.3\arcmin~from the centre of the system ($\simeq 1.8 r_h$, see Sect.~\ref{sec:structural_params}).
The absolute magnitude of the HB is a valuable distance indicator, and we used it as the strongest constraint on the distance modulus. Through a visual isochrone fitting, we re-determined the parameters that provide the best match to the observed CMD: $\log (t/yrs) \simeq 10.12$~dex; $[Fe/H] \simeq$ -2.2 dex and $m - M$ = 20.8 $\pm$ 0.2 mag. This is the red solid line superimposed on all panels in the second row of Fig.~\ref{fig:kids_density_maps}.\par
The resulting distance modulus corresponds to a physical distance of approximately 150 kpc, well within the range of Galactic satellites. The green isochrone overlaid on the panel $a$ of the second row of Fig.~\ref{fig:kids_density_maps} deviates from the best-matching isochrone only in terms of age. Specifically, it has ${\rm log~t} = 10.00$~dex, showing that decreasing the age of the stellar model by $\sim \pm 2-3$ Gyr does not significantly affect the absolute magnitude of the HB (hence our distance estimate), as expected.\par 
Finally, the panel $d$ of the second row of Fig.~\ref{fig:kids_density_maps} presents the CMD of Sextans~II, including all stars confined within 4 $\times~r_h$. 
This larger region enables a more comprehensive search for additional HB stars that might be associated with the system, albeit at the cost of higher contamination from MW stars within the CMD. We identify a potential fifth HB member at $\sim$15.9\arcmin~from the galaxy centre, corresponding to 3.4 $\times r_h$. This star has a colour of -0.09 mag, similar to the other three HB stars, and indeed it is barely discernible in the main panel. To enhance the visibility of these four closely colour-matched HB stars, we provide a zoomed-in view of the HB evolutionary phase in an inset at the top of the panel $d$, confined to a narrow range of colours encompassing those of the four stars.\par

\subsection{Structural parameters}
\label{sec:structural_params}

To estimate the structural parameters, such as the centre, half-light radius, ellipticity, and position angle of the newfound stellar system, we adopted the maximum likelihood approach presented in \citet{Martin-2008}. 
In particular, \citet{Martin-2008} model the two-dimensional position of the stars in the sky in the surroundings of the galaxy with the combination of the exponential profile of the galaxy and of a uniform background.\par
    


\begin{figure}
    \centering
    \includegraphics[width=.5\textwidth]{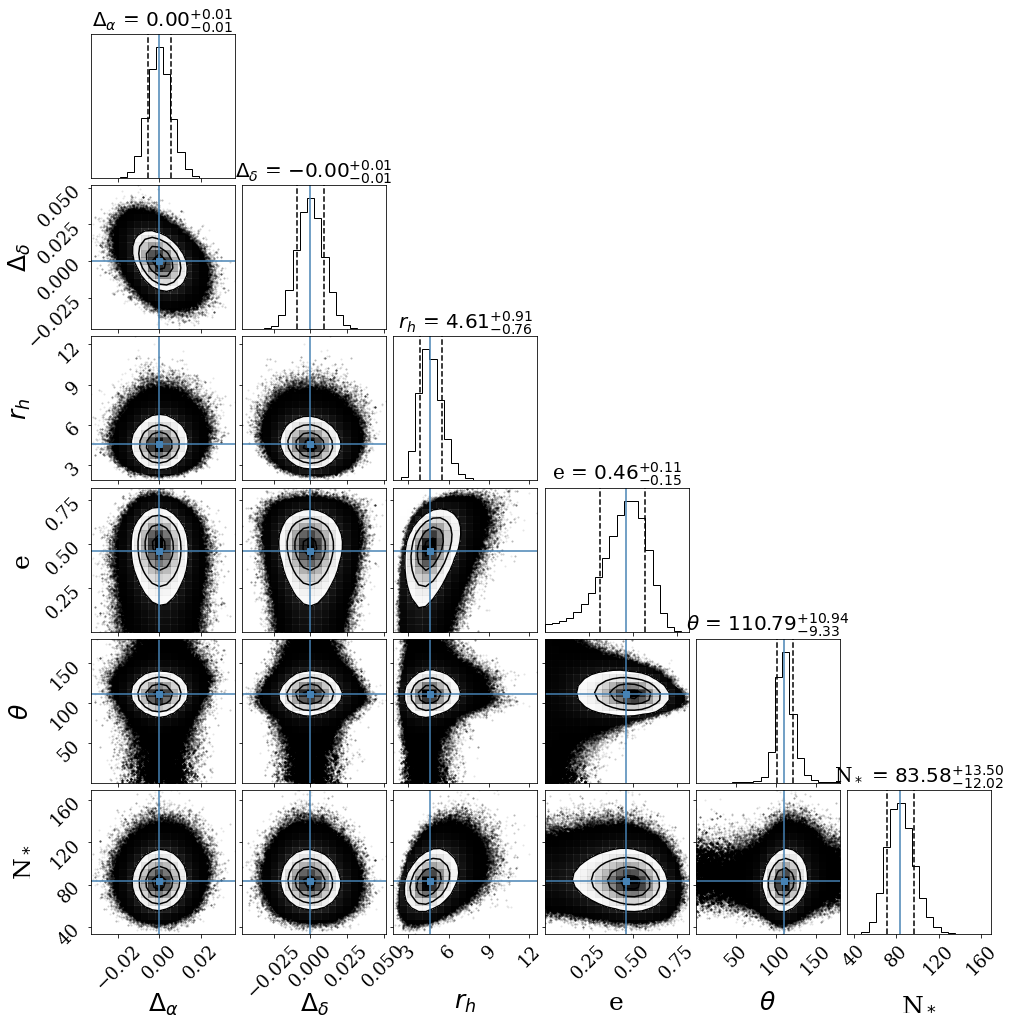}\\
    \caption{Posterior probability distributions for each of the six free parameters of the model, derived from the MCMC approach described in Section~\ref{sec:structural_params}.}
    \label{fig:mcmc}
\end{figure}

We adopted a Markov Chain Monte Carlo (MCMC) technique to contemporaneously estimate the free parameters of our model: $x_0; y_0; r_h; e; \theta; N_{\rm stars}$, employing only stars located within the best-fit isochrone mask.
Our MCMC implementation utilized 1000 different walkers, each executing a chain composed of 10000 steps. We discarded the first 1000 steps as burn-in, resulting in a robust estimation of the model parameters. The MCMC analysis was conducted using the \emph{emcee} \citep{emcee} Python library. The best-fit parameters and their uncertainties were determined from the median, 16th, and 84th quantiles of the posterior distributions.
In Fig.~\ref{fig:mcmc}, we present a corner plot displaying the posterior distribution of the six free parameters, while Table~\ref{tab:galaxy_params} lists the median values and uncertainties of these parameters.\par
\begin{figure*}
    \centering
    \includegraphics[width=.45\textwidth]{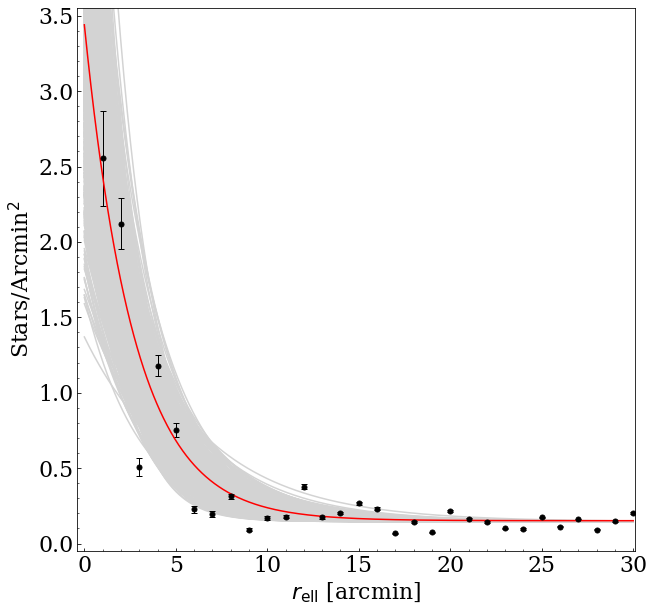}
    \includegraphics[width=.45\textwidth]{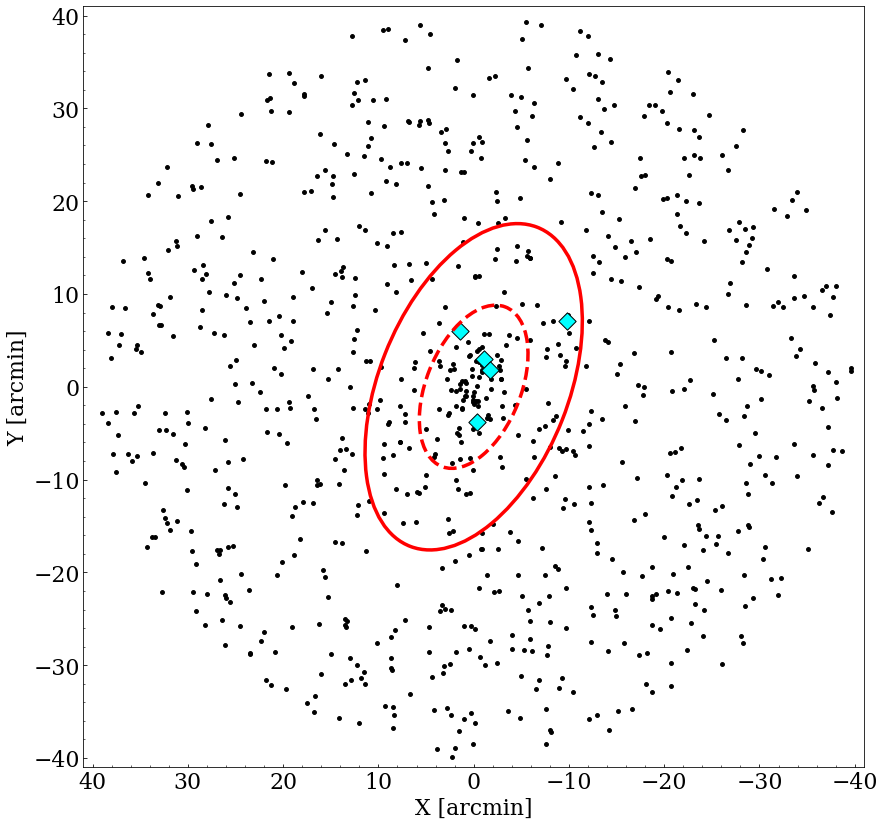}
    \caption{\emph{Left}: Radial Density Profile of KiDS-UFD-1. $r_{\rm ell}$ is the elliptical radius defined in eq.4 in \citet{Martin-2008}.
    The red solid line refers to the best-fit model with parameters derived through the MCMC approach described in Section~\ref{sec:structural_params}, while gray lines represent 1000 random extractions from the posterior distributions of the free parameters of the model. \emph{Right}: Relative position of all stars filtered by the best-fit isochrone with respect to the galaxy center. The red dashed and solid ellipses have ellipticity and position angle derived from the MCMC procedure and semi-major axes of $2\times r_h$ and $4\times r_h$, respectively. Cyan diamonds indicate the relative position of the five HB stars.}
    \label{fig:2D_distr}
\end{figure*}

 
The left panel of Fig.~\ref{fig:2D_distr} illustrates the radial density profile (RDP) of the system, where $r$ represents the elliptical radius as defined in eq.~4 in \citet{Martin-2008}. 
The red solid line denotes the best fit of the model for the two-dimensional position of the stars in the sky derived by employing the median values of the six parameters obtained through MCMC analysis. It is interesting to note that, both the derived half-light radius $r_h = 194^{+61}_{-45}$ pc (assuming $D_{\odot} = 145$~kpc), and ellipticity $e=0.46^{+0.11}_{-0.15}$ are well within the range typical for local UFDs \citep[see the latest version of the][catalogue]{McConnachie2012}.
The right panel of Fig.~\ref{fig:2D_distr} displays the 2-D relative position of the filtered stars with respect to the galaxy centre, along with ellipses aligned according to the position angle determined by the MCMC procedure. These ellipses are defined by semi-major axes of $2r_h$ (dashed ellipse) and $4r_h$ (solid ellipse). Additionally, we indicate the relative position of the five stars in the HB evolutionary stage as cyan diamonds. As already stated, four HB stars are situated within $2r_h$, while the fifth one is at almost 4$r_h$.

\begin{table}[]
    \caption{Properties of Sextans~II (KiDS-UFD-1)}
    \label{tab:galaxy_params}
    \centering
\begin{tabular}{lc}
\hline\hline
  Quantity   &  Value\\
    
    \hline
      R.A. [J2000] & 156.4399 deg\\ 
      Dec [J2000] & -0.6400 deg\\
      t & $\ga 10$ Gyr \\
      ${\rm [Fe/H]}$ & $\la -1.5$ dex\\
      E(B-V) & 0.056 mag\\
      ${\rm (m - M)_0}$  & 20.8 $\pm$ 0.2 mag\\
      D  & 145$^{+14}_{-13}$ kpc\\
      $r_h$ [arcmin] & 4.6$^{+0.9}_{-0.8}$ arcmin\\
      $r_h$ [pc] & 194$^{+61}_{-45}$ pc\\
      $e$ & 0.46$^{+0.11}_{-0.15}$\\
      $\theta$ & 111$^{+11}_{-9}$ deg\\ 
      $M_*$ & 4900$^{+1300}_{-1200}$ $M_{\odot}$\\
      $M_r$  & -4.1 $^{+0.5}_{-0.4}$ mag\\ 
      $M_g$  & -3.5$^{+0.4}_{-0.3}$ mag\\
      $M_V$  & -3.9$\pm 0.4$ mag\\
      
    \hline
    \end{tabular}

\end{table}

\subsection{Integrated magnitude and stellar mass}

In this section, we outline the procedure employed to estimate the absolute magnitude in the V-band for the purpose of comparing the properties of Sextans~II with other recently discovered UFDs. 
Our approach follows the methodology reported in \citet{Martin-2008}, which is based on Monte Carlo sampling of a synthetic CMD representative of Sextans~II.
Specifically, we built a synthetic stellar population with $\log (t/yrs) = 10.12$~dex; [Fe/H] $\sim$-2.2 dex, placed at a distance modulus 20.8 mag. We considered a total of $N_{\rm stars} = 52 \pm 12$, that is the value obtained from the MCMC approach by adopting a cut at $r = 24$~mag, which is a conservative threshold adopted to mitigate potential issues of completeness in the central regions of the galaxy that might bias the estimation of its absolute magnitude.
We generated 1000 realizations of the CMD, populating it with a \citet{Kroupa2002} initial mass function (IMF), and subsequently computed the average absolute magnitude in the $g$ and $r$ SDSS filters, as well as in the Johnson $V$ band\footnote{To convert the $g$ SDSS filter magnitude into the Johnson $V$ band magnitude we applied the conversion relationship described in \citet{Jester-2005}.}. Additionally, we determined the stellar mass of the galaxy. The median values, along with the 16th and 84th quantiles, were adopted to represent the best values and associated uncertainties, respectively. These results are presented in the final rows of Table~\ref{tab:galaxy_params}.\par
\begin{figure}
    \centering
    \includegraphics[width=.45\textwidth]{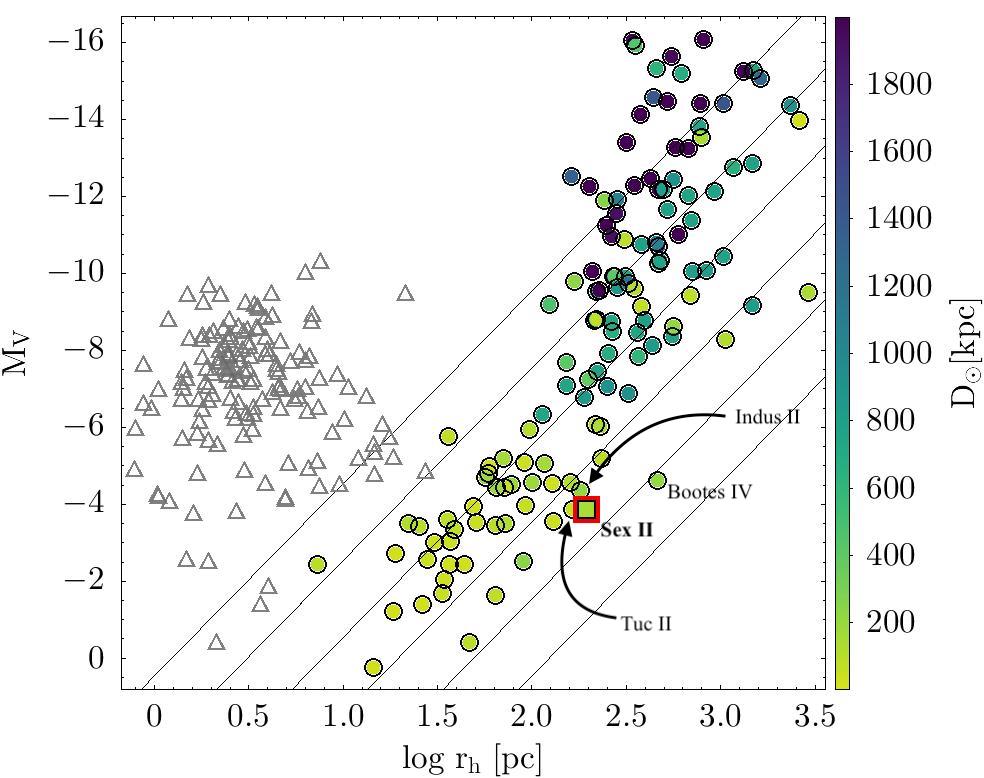}
    \caption{Absolute integrated $V-$band magnitude  as a function of the log of the half-light radius for local low-luminosity stellar systems.
    Galactic globular clusters, represented as open grey triangles, are from the 2010 version of the \citet{Harris1996} catalogue. 
    Confirmed and candidate dwarf galaxies, represented as filled circles color-coded according to their heliocentric distance, are taken from the 2021 version of the catalog by \citet{McConnachie2012}. The same colour-coding is applied to the filled square marking the position of Sextans~II, that is also highlighted with a red contour. The stellar systems with $M_V$ and $r_h$ most similar to Sextans~II have been labelled. The diagonal lines are loci of constant surface brightness, from at $\mu = 24.0$~mag arcsec$^{-2}$ (top) down to $\mu = 34.0$~mag arcsec$^{-2}$ (bottom), in steps of $\mu = 2.0$~mag arcsec$^{-2}$.}
    \label{fig:Mv_vs_rh}
\end{figure}

 
Figure~\ref{fig:Mv_vs_rh} depicts the $M_V$-$r_h$ plane for galaxies and candidate galaxies reported in catalog by \citet{McConnachie2012} (version 2021), as well as for Galactic Globular Clusters included in the catalog by \citet{Harris1996} (version 2010).
The position of Sextans~II within this plot clearly establishes it as one of the UFDs with the lowest surface brightness discovered to date, although it is important to note that some extremely faint low-surface brightness galaxies identified in recent years may not be represented in this figure.


\section{Summary and conclusions}

In this letter we report on the discovery of an old and metal-poor stellar system located in the outskirts of the Milky Way, Sextans~II (KiDS-UFD-1).
This discovery resulted from an extensive search for low-surface brightness stellar systems in the MW halo within the KiDS survey catalogues.
The KiDS detection, limited by photometric depth to the RGB and HB of the system, was confirmed with deeper HSC-SSP photometry
which clearly unveiled the old MS of Sextans~II down to about 1 mag below the MSTO.
The isochrone fitting of the CMD revealed that the newly discovered system is ancient (t $\geq$ 10 Gyr) and metal-poor ([Fe/H] $\leq$ -1.5), and is located at a distance of $D_{\odot} \simeq 145$~kpc from the Sun and $R_{GC}\simeq 128.8$~kpc from the center of the Galaxy.
The structural parameters of Sextans~II
are  typical of UFDs in the Local Group, strongly suggesting that the new system is a faint spheroidal satellite of the MW at the low end of the surface brightness distribution of local dwarfs. The most plausible alternative hypothesis is a disrupting globular cluster. However, even the most extended known globular clusters in the MW have $r_h\le 30.0$~pc, more than five times smaller than what we determined for Sextans~II, hence the UDF hypothesis is, by far, the most likely.
The final word on the nature of the system can be provided only by a proper spectroscopic follow up of a reasonable sample of member stars, that may be challenging, given the magnitude range spanned by candidate RGB members\footnote{In the recently published FASHI catalogue of HI sources \citep{Zhang-2023} we found a source at cz=10443.8 km/s (FASHI 20230004672, estimated distance $\sim 140$~Mpc) that coincides in the sky with Sextans~II, has a characteristic size ($3.9\arcmin$) comparable with the $r_h$ of Sextans~II, and a mass compatible with a galaxy group. 
While we are confident that our colour cuts get rid of most of the contamination by unresolved galaxies, only spectroscopic follow-up will ascertain if there is significant contamination of Sextans~II from galaxies possibly associated with this distant HI source.}. The chemical composition can reveal the presence of a metallicity spread, while the kinematics can reveal the amount of 
non-baryonic dark matter required to keep the system at the dynamical equilibrium, both features that are generally used to discriminate between dwarf galaxies and star clusters \citep{willman2012}. \par
The research of unknown stellar systems is still ongoing, and Sextans~II might not be the only newfound stellar system detected in the KiDS DR4 catalog. Moreover this work underscores the importance of extending this research to the forthcoming fifth release of the KiDS survey, that will provide multi-band photometric data covering the entire 1350 square degrees of the KiDS footprint.
This discovery serves as a reminder that the census of the Milky Way's satellite galaxy population is far from complete. The upcoming wide and deep surveys, such as the Legacy Survey of Space and Time (LSST) set to be conducted at the imminent Vera Rubin Observatory, will likely contribute significantly to expanding our knowledge of Milky Way satellites, much like the impactful contributions of the SDSS and PanSTARRS surveys before it.


\begin{acknowledgements}
Based on observations made with ESO Telescopes at the La Silla Paranal Observatory under programme IDs 177.A-3016, 177.A-3017, 177.A-3018 and 179.A-2004, and on data products produced by the KiDS consortium. The KiDS production team acknowledges support from: Deutsche Forschungsgemeinschaft, ERC, NOVA and NWO-M grants; Target; the University of Padova, and the University Federico II (Naples).\par
The Hyper Suprime-Cam (HSC) collaboration includes the astronomical communities of Japan and Taiwan, and Princeton University. The HSC instrumentation and software were developed by the National Astronomical Observatory of Japan (NAOJ), the Kavli Institute for the Physics and Mathematics of the Universe (Kavli IPMU), the University of Tokyo, the High Energy Accelerator Research Organization (KEK), the Academia Sinica Institute for Astronomy and Astrophysics in Taiwan (ASIAA), and Princeton University. Funding was contributed by the FIRST program from the Japanese Cabinet Office, the Ministry of Education, Culture, Sports, Science and Technology (MEXT), the Japan Society for the Promotion of Science (JSPS), Japan Science and Technology Agency (JST), the Toray Science Foundation, NAOJ, Kavli IPMU, KEK, ASIAA, and Princeton University. 
This paper is based on data collected at the Subaru Telescope and retrieved from the HSC data archive system, which is operated by the Subaru Telescope and Astronomy Data Center (ADC) at NAOJ. Data analysis was in part carried out with the cooperation of Center for Computational Astrophysics (CfCA), NAOJ. We are honored and grateful for the opportunity of observing the Universe from Maunakea, which has the cultural, historical and natural significance in Hawaii.\par
M.G. acknowledges the INAF AstroFIt grant 1.05.11.
      C.T. acknowledges the INAF grant 2022 LEMON. JTAdJ is supported by the Netherlands Organisation for Scientific Research (NWO) through grant 621.016.402.
      H. Hildebrandt is supported by a DFG Heisenberg grant (Hi 1495/5-1), the DFG Collaborative Research Center SFB1491, as well as an ERC Consolidator Grant (No. 770935).
      KK acknowledges support from the Royal Society and Imperial College.
      SZ acknowledges the support from the Deutsche Forschungsgemeinschaft (DFG) SFB1491.
      NRN acknowledges financial support from the National Science Foundation of China, Research Fund for Excellent International Scholars (grant n. 12150710511), and from the research grant from China Manned Space Project n. CMS-CSST-2021-A01.
\end{acknowledgements}

%
%

\bibliographystyle{aa} 
\bibliography{Dwarf} 

\end{document}